\documentclass[10pt,letterpaper]{article}
\usepackage[top=0.85in,footskip=0.75in,marginparwidth=2in]{geometry}

\usepackage[utf8]{inputenc}

\usepackage{cite}

\usepackage{nameref,hyperref}

\usepackage[right]{lineno}

\usepackage{microtype}
\DisableLigatures[f]{encoding = *, family = * }

\raggedright
\usepackage{changepage}

\usepackage[aboveskip=1pt,labelfont=bf,labelsep=period,singlelinecheck=off]{caption}

\makeatletter
\renewcommand{\@biblabel}[1]{\quad#1.}
\makeatother

\usepackage{lastpage,fancyhdr,graphicx}
\usepackage{epstopdf}
\fancyheadoffset[L]{2.0in}
\fancyfootoffset[L]{2.0in}

\usepackage{color}

\definecolor{Gray}{gray}{.25}

\usepackage{graphicx}
\usepackage[export]{adjustbox}

\usepackage{sidecap}

\usepackage{wrapfig}
\usepackage[pscoord]{eso-pic}
\usepackage[fulladjust]{marginnote}
\reversemarginpar

\begin{document}
\vspace*{0.35in}

\begin{flushleft}
{\Large
\textbf\newline{IMPACT CRATER MORPHOLOGY AND THE STRUCTURE OF EUROPA'S ICE SHELL}
}
\newline
{\Large
\textbf\newline{(paper accepted on 8 Nov 2017 for publication in JGR - Planets, doi: 10.1002/2017JE005456)}
}
\newline
\\
Elizabeth A. Silber\textsuperscript{1,*},
Brandon C. Johnson\textsuperscript{1}
\\
\bigskip
\textsuperscript{1}Department of Earth, Environmental and Planetary Sciences, Brown University, Providence, RI, USA, 02912
\\
\bigskip
*elizabeth\textunderscore silber@brown.edu
\\
\bigskip
\textbf{Citation: Silber, E. A. and B. C. Johnson (2017) Impact crater morphology and the structure of Europa's ice shell, JGR - Planets, doi: 10.1002/2017JE005456} 

\end{flushleft}

\section*{Abstract}
\justify
We performed numerical simulations of impact crater formation on Europa to infer the thickness and structure of its ice shell. The simulations were performed using iSALE to test both the conductive ice shell over ocean and the conductive lid over warm convective ice scenarios for a variety of conditions. The modeled crater depth-diameter is strongly dependent on thermal gradient and temperature of the warm convective ice. Our results indicate that both a fully conductive (thin) shell and a conductive-convective (thick) shell can reproduce the observed crater depth-diameter and morphologies. For the conductive ice shell over ocean, the best fit is an approximately 8 km thick conductive ice shell. Depending on the temperature (255 – 265 K) and therefore strength of warm convective ice, the thickness of the conductive ice lid is estimated at 5 – 7 km. If central features within the crater, such as pits and domes, form during crater collapse, our simulations are in better agreement with the fully conductive shell (thin shell). If central features form well after the impact, however, our simulations suggest a conductive-convective shell (thick shell) is more likely. Although our study does not provide firm conclusion regarding the thickness of Europa’s ice shell, our work indicates that Valhalla-class multiring basins on Europa may provide robust constraints on the thickness of Europa’s ice shell. 


\section*{1. Introduction}
Geophysical observations of Jupiter’s moon Europa suggest that this icy world hosts a global subsurface ocean [e.g., Khurana et al., 1998; Kivelson et al., 1999; 2000]. Europa’s ice shell has been estimated to be a few [e.g., Hoppa, 1999; Greenberg et al., 2000] to a few tens of kilometers thick [e.g., Pappalardo et al., 1999; Prockter and Pappalardo, 2000; Moore et al., 2001]. The thickness and structure of the ice shell bear immense relevance to astrobiological, geophysical and planetary body evolution considerations, as well as future space mission planning. On one hand, a thin shell over ocean would enable impact penetration [e.g., Cox et al., 2008], formation of chaos regions, heat exchange via conduction, and nutrient exchange via direct cracks [e.g., Greenberg et al., 2000]. On the other hand, a thick shell, composed of a conductive stagnant lid and warm convecting ice underneath, would enable tidal heating dissipation [e.g., McKinnon, 1999; Nimmo and Manga, 2009], while diapirs initiated by thermal convection could be a mode for nutrient exchange [e.g., Ruiz et al., 2007] and formation of chaos regions [e.g., Schmidt et al., 2011]. Despite recent developments in remote sensing [e.g., Schenk, 2002] and numerical modeling [e.g., Bray et al., 2014; Cox and Bauer, 2015], large uncertainties regarding the thermal structure and thickness of Europa’s ice shell remain. 
\par Impact craters produced by hypervelocity collisions are one of the most ubiquitous geological features on solid planetary surfaces. The final size and morphology of an impact crater depend on both the projectile (e.g. size, velocity) and target properties (e.g. rock, ice, thermal gradient) [e.g., Melosh, 1989]. Impact craters exhibit a progression in morphologies as a function of increasing diameter. On rocky bodies, simple, bowl shaped crater morphologies transition to complex craters with flat floors and central peaks, then peak-ring basins, and finally multiring basins [e.g., Pike, 1980; Melosh, 1989]. During an impact, a bowl shaped transient crater is excavated and subsequently succumbs to gravitational collapse or modification. Crater modification, and therefore the resulting crater dimensions and morphology, are highly sensitive to target properties, such as heat flow, layering, and the presence of underlying ocean [e.g., Collins et al., 2004; Ivanov et al., 2010; Senft and Stewart, 2011; Johnson et al., 2016a]. Hypervelocity impacts generate tremendous amounts of energy, melting and vaporizing both the projectile and target. Ice is brittle and behaves like typical terrestrial rock at Europa’s surface temperature of 100 K [Ojakangas and Stevenson, 1989]. However, considering that ice melts and deforms at temperatures substantially lower than that for rock, icy bodies will be even more susceptible to expressing the conditions at depth through crater morphologies [e.g., Schenk, 2002]. Among tools used to estimate the thickness of Europa’s ice shell, analysis of impact craters remains the most cost-effective probe of ice shell thickness and likely conditions at depth [e.g., Moore et al., 1998; Schenk, 2002; Bray et al., 2008, 2014; Cox and Bauer, 2015].
\par Observations of impact craters, characterized as fresh, unmodified and unrelaxed, by Schenk [2002], on the icy Galilean moons reveal the depth-diameter (d-D) relationship that exhibits three distinct transitions. On Europa, these transitions are: simple-to-complex transition (I), anomalous crater dimensions and morphologies (e.g. central pits and domes) with crater depths that decrease as crater diameter increases (II), and abrupt transition from modified central peak to shallow craters with multiring morphologies (III) [Schenk, 2002] (Figure 1). Decreasing crater depths with increasing diameters, transitions II and III, are unique to the icy Galilean moons, occurring on Europa at D ~4 km and ~8 km, respectively and larger diameters on Ganymede and Callisto. Schenk [2002] argued that these transitions on Europa are indicative of warm convecting ice at depths of 7-8 km and a liquid ocean at 19-25 km, respectively [Schenk, 2002]. Figure 2 shows various crater morphologies of Europan craters ranging from 1.3 km to 23 km in diameter.
\par Numerical modeling by Turtle and Pierazzo [2001] demonstrated that Europa’s ice shell must be thicker than 3–4 km, otherwise the zone of material melted by the impact would intersect the underlying ocean and thus inhibit formation of craters with central peaks, which require significant material strength. Recent developments and improvements in numerical models have opened the door for more sophisticated studies. Bray et al. [2014] estimated the ice shell thickness at 7 km, while simultaneously matching the observed crater morphologies (e.g. summit pit and dome craters) and depth-diameter relationships [Schenk, 2002] (Figure 1). Cox and Bauer [2015] placed Europa’s icy shell thickness at 10 km by numerically probing the effects of crust breaching and matching the observed crater depth-diameter relationships (Figure 1), regardless of observed morphology. Both studies [Bray et al. 2014; Cox and Bauer, 2015] considered only a conductive ice shell over ocean scenario. 
\par We modeled formation of impact craters to probe its internal structure. Our study differs from previous works in that we consider both a fully conductive ice shell over ocean and a conductive ice shell over warm convective ice to attempt to discern boundary conditions at the interface between the ice and the underlying ocean (Figure 3). In the latter scenario, it is assumed that a conductive ice lid and warm convective ice overlie the global ocean. We aim to elucidate the following questions regarding the anomalous morphology of Europan craters: Is the conductive ice shell over ocean the only plausible scenario? Could the conductive lid over warm convective ice (thick shell) scenario reproduce the observed depth-diameter and crater morphologies on Europa as well as the fully conductive ice shell (thin shell)? If yes, what are some probable parameters associated with the structure of the shell, specifically thickness of the conductive lid and the temperature of warm convective ice? 
\par Our paper is organized into several sections. The methodology and model setups are described in Section 2. Section 3 outlines our results, and is further divided into subsections that discuss the modeled crater depth-diameter for conductive (thin) ice shell (3.1.1) and conductive-convective (thick) ice shell (3.1.2), and the comparison to observed crater morphologies (3.2) in the context of both simple-to-complex transition (3.2.1) and anomalous morphologies (3.2.2). Section 4 presents the discussion about the numerical modeling considerations relevant to formation of Europan crater morphologies through hydrocode simulations (4.1), and the effect of strength of ice on crater morphology (4.2). In this section we further discuss our findings in the context of pits and domes formation (4.3), and the thin versus thick ice shell, followed by implications for future studies (4.4). The conclusions are listed in Section 5. 

\section*{2. Methods}
We modeled the formation of impact craters on Europa using iSALE-2D, a multi-material, multi-rheology shock physics code [Melosh et al., 1992; Ivanov et al., 1997; Collins et al., 2004; Wünnemann et al., 2006], which is based on the SALE hydrocode solution algorithm [Amsden et al., 1980]. Due to the axial symmetry of our models, only vertical impacts were considered. Although the average impact velocity ($v_i$) for Europa is 26 km/s [Zahnle et al., 2003], we use $v_i$ = 15 km/s to reduce the simulation time and to maintain the consistency with previous work [Bray et al., 2014; Cox and Bauer, 2015]. The ocean layer was represented by the ANEOS [Thompson and Lauson, 1972] and ice Ih by Tillotson equations of state (EOS) [Tillotson, 1962; Ivanov et al., 2002]. 
The strength and damage models for ice shell, and the parameters for the Block Model of acoustic fluidization [Melosh, 1979] correspond to Bray et al. [2014] and are listed in Table 1 and Table 2, respectively. In the Block Model, the parameters controlling the degree to which the target is weakened during cratering are decay time and limiting viscosity of the fluidized target. Bray et al. [2014] estimated these empirically by performing a broad parameter study for uniform non-layered ice to match the morphometry of impact craters on Ganymede. Since large craters (D $>$ 25 km) are more susceptible to rheological changes that occur with depth [Schenk 2002], the parameters were determined for small impact craters small (D = 4 – 25 km). 
The grid resolution was set to 10 cells per projectile radius, as that offers optimal computing time, and a reasonably small error, while retaining the temporal evolution and morphology of the resulting crater [e.g., Elbeshausen et al., 2009; Silber et al., 2017 and references therein]. The impactor radii (Ri) also correspond to Bray et al. [2014] and range from 26 m to 405 m (Table 2). Similar impactor sizes were also employed by Cox and Bauer [2015].  
\par We considered a full viscoelastic-plastic ice rheology to account for any viscous contribution to material deformation. A viscoelastic-plastic rheology for rocky mantle material was first introduced in iSALE by Dirk Elbeshausen and is available as part of the iSALE-Dellen release (https://dx.doi.org/10.6084/m9.figshare.3473690). Parameters appropriate for ice were subsequently implemented by Johnson et al. [2016b]. This implementation adds a viscous term to iSALE elastic-plastic rheology, and permits for modeling and full treatment of an ice shell with warm convecting ice. For more information, see the supplementary information of Johnson et al. 2016b].  A simple comparison of Maxwell time to crater collapse time indicates that viscoelastic-plastic rheology of ice might have important effect for craters $>$ $\sim$~10 km in diameter. 
\par Our study consists of two components. First, the aim was to model the fully conductive ice shell over the liquid ocean (Figure 3a) (as also done by Bray et al. [2014] and Cox and Bauer [2015]) for comparison to our simulation that include warm convecting ice and to investigate the influence of a viscoelastic-plastic ice rheology. To this end, we ran several suites of simulations. We modeled the fully conductive shell with thickness 7, 8 and 9 km using the viscoelastic-plastic ice rheology implementation. For comparison purposes, we also ran one simulation suite for the 7 km ice shell without the viscoelastic-plastic ice rheology implementation. Estimates of Europa’s surface heat flow range from approximately 50 mW/m\textsuperscript{2} [Ojakangas and Stevenson, 1989; Melosh et al., 2004] up to $\sim$~100 mWm\textsuperscript{-2} [Ruiz and Tejero, 2000]. A purely conductive layer is produced assuming a linear thermal gradient (dT/dz) corresponding to the surface temperature of 100 K [Ojakangas and Stevenson, 1989] and temperature of 273 K for the underlying ocean. Thus, for a 7 km conductive ice shell, the heat flow (q) is $\sim$~74 mWm\textsuperscript{-2} and thermal conductivity (k) is $\sim$~3 J/m/s [Bray et al., 2014]. Values of thermal gradient for various ice shell thicknesses are given in Table 3.
\par Second, we modeled conductive-convective layering, where conductive ice (a conductive lid) with a steep thermal gradient (Table 3) overlays a region of convective warm ice that has an adiabat of 0.05 K/km. Schenk [2002] estimated the depth of the ocean at 19 - 25 km and noted that craters beyond Transition III would exhibit features indicative of the rheological changes (e.g., multiring structures). In comparison, the transient crater depth ($d_{tr}$) for largest impactors (Ri = 405 m) considered in this study is approximately 6 km, which is a factor of 3 – 4 less than the estimated depth of the ocean [Schenk, 2002]. Therefore, since the presence of the ocean at depths 19 - 25 km is not expected to make an appreciable contribution at crater sizes investigated in this study ($d_{tr}$ $<$ 6 km, D $<$ 23 km) (also see Figure 4a in Cox and Bauer [2015]), we do not explicitly model the ocean layer in our conductive-convective simulations. Nevertheless, it is implicitly assumed that the global ocean indeed exists in all setups (Figure 3b). To account for a range of possible scenarios, we varied both the conductive lid thickness (4 – 7 km) and the temperature of the warm convective ice (255 K and 265 K) [e.g., Pappalardo et al., 1999; McKinnon, 1999]. Examples of strength-depth and temperature-depth profiles are shown in Figure 4. The strength (Y) of ice depends on layering (conductive ice over ocean or conductive ice overlying warm convective ice) and temperature (Figure 4a). In the conductive region, a thicker conductive layer corresponds to higher peak strength. In the convective (adiabatic) region of the shell, higher temperature leads to lower strength (Y = 0 for liquid).
\par The crater diameters were measured rim to rim, and crater depths from the highest point on the rim to the lowest point on the crater floor. The uncertainty in depth measurements was set to two grid cells. Once cell corresponds to the physical size of Ri/10. The uncertainty in crater diameter ($D_{err}$) measurement was determined in the following way. Since the highest point on the crater rim can be anywhere from one to several cells wide, depending on a simulation, we first found the median (R), which also serves as the measurement of the crater radius (D = 2R). Then, we obtained the intersecting point ($R_{wall}$) of the crater wall and the pre-impact surface (Figure 5). The uncertainty in crater radius is expressed as $R_{err}$ = 0.25(R - $R_{wall}$), and in overall diameter as $D_{err}$ = 2$R_{err}$. The uncertainty in crater depth is given as one grid cell in each direction (+/-), corresponding to the physical size of Ri/10 and thus ranging from 2.6 m (Ri = 26 m) to 40.05 (Ri = 405 m). 
\par Finally, since iSALE is not suitable for studying post-impact relaxation or small scale deformations over long time scales (e.g., after the crater collapse is over), caution should be given when determining simulation run times. Extended simulation run times result in reflections, which coupled with numerical diffusion, may give the appearance of shallowing craters. A couple of approaches to determine when a simulation is over can be applied. One is to qualitatively establish when all the motion stops from plotted time series. The other, more robust, is to plot up crater profiles (depth versus diameter) across several time frames before and after crater collapse is deemed to have finished, and find the point at which crater depth, diameter and morphology no longer undergo a notable change. Examples of this are shown in Figure 6. In Figure 6a, the reported simulation end time ($t_{end}$) is 215 s. The three subsequent time frames (t = 220, 225 and 230 s) show that crater depth, diameter and morphology no longer undergo a notable change. The ‘steps’ (e.g., on crater floor and crater rim), represent discrete cells in the simulation grid. Similarly, in Figure 6b, the crater profile at t = 400 s is nearly identical to that at $t_{end}$. The top of the crater rim is narrowed laterally by only one cell on each end, and the central peak is also reduced by one cell laterally. However, this makes no difference in either the measured crater depth or diameter, because the methodology to measure the crater dimensions described earlier in this section is robust. For example, both R and $R_{err}$ remain unchanged from tend to t = 400 s. The uncertainty in crater depth, corresponding to a single cell in each direction, also accounts for any minute variations. 

\section*{3. Results}
Figure 7 shows the formation of anomalously shallow craters by a 320 meter radius impactor for two possible ice shell structures. The left panel represents a 6 km thick conductive lid over warm convective ice at 265 K and the right panel is for an 8 km thick ice shell over ocean. As we will show later, these conditions represent our best fit for reproducing observed crater depths and diameters. The collapse of the transient crater results in a large central uplift (Figure 7a), which subsequently collapses (Figure 7b). The central uplift is much larger and reaches more than twice as high for the convective ice case as compared to the 8 km conductive ice shell (Figure 7). In the 8 km shell, the presence of the underlying ocean leads to pronounced crater floor uplift seen at 240 s (Figure 7b), which is retained in the final crater morphology (Figure 7c). While there is no significant crater floor uplift retention in the convective ice scenario, the final crater is anomalously shallow (Figure 7c). Regardless of the differences in the final morphology at the center of the crater, both simulations yield a crater that is 0.4 km deep and the diameters are similar (D = 18 km for the 6 km conductive lid over convective ice and D = 17 km for the 8 km conductive shell over ocean). We will discuss these differences in Section 3.2.
\subsection*{3.1 Modeled Crater Depth-Diameter}
In Figure 8 we compare the observed crater depth-diameter (d-D) [Schenk, 2002] to our simulation results. The conductive ice shell over ocean (Figure 8a), and the 255 K (Figure 8b) and 265 K (Figure 8c) warm convective ice scenarios are plotted in separate panels for clarity. The modeled crater diameters and depths, along with their uncertainties in measurement, are listed in Tables S1 – S3. 
Although we ran simulations for impactors up to and including Ri = 405 m for all scenarios, in several instances, specifically for high thermal gradient (dT/dz = 28.8 – 43.3 K/km) and large impacts (Ri = 320 – 405 m), it was not possible to reliably measure the crater diameter or depth due to warm material overflowing over the initial crater rim, thus skewing the post-impact dimensions (denoted with asterisk in Tables S1 – S3). We describe this in more detail in Section 3.3. Triangles in Figure 8(b,c) represent craters for which it was not possible to reliably measure d-D due to warm material overflow, and half-filled points represent the craters for which it was possible to determine the location of the crater rim before it was engulfed by the warm material (see Section 3.3). The specific cases will be referred to further in this section.
While crater diameters produced by impactors of a given size are relatively insensitive to changes in pre-impact thermal structure considered here, the depths are significantly different and strongly dependent on the temperature gradient. The implementation of viscoelastic-plastic rheology produces relatively small differences in crater depth and diameters for the crater size range investigated here (D $<$ 23 km), and no notable differences in crater morphology. 
\subsection*{3.1.1 Conductive Ice Shell}
In the ice shell over ocean scenario (Figure 8a), for crater sizes up to ~8 km in diameter (Ri = 150 m), all models produce very similar d-D, which is also consistent with observations. At smaller crater sizes (Ri = 107 m, D $\sim$~ 5.7 km), the crater depths and diameters for all three shell thicknesses converge, implying that around and below this threshold, impact craters are insensitive to the shell thickness and thermal gradient. At diameters larger than ~ 8 km, the 7 km ice shell results in craters that are on the shallow end of the observed d-D, with significant divergence and a steep drop off in crater depth beyond D ~15 km. For the simulations with a 9 km shell, the 11 km crater (Ri = 230 m) is too deep, while the craters in larger size range are within the observed d-D bounds. The 8 km ice shell produces the best fit to the d-D of observed craters. An animation of the best fit simulation for Ri = 230 m is in Supplementary Material (Animation S1). This result is broadly consistent with the estimates derived from earlier numerical studies [Bray et al. 2014; Cox and Bauer 2015]. 
\subsection*{3.1.2 Conductive Ice Lid Over Warm Convective Ice}
In the 255 K warm convective ice scenario (Figure 8b), the 7 km conductive lid results in a flat d-D, with the roll-off nearly absent, and with craters that are too deep. Both the 5 km and 6 km lid over warm convective ice appear more consistent with the observations; however, the rollover into Transition II [Schenk, 2002] is not well defined for a 6 km case. The 4 km conductive lid results in very shallow craters, especially beyond D $\sim$~ 8 km. The d-D for Ri = 320 and 405 m was not plotted, because it was not possible to measure the crater diameter (see Section 3.2). The 5 km conductive lid over warm convective ice is well within the observed d-D, although the largest crater is among the deepest observed craters for the given diameter (D $\sim$~ 21 km). The warm convective ice at 265 K (Figure 8c) with the 5 km conductive lid yields relatively shallow craters, with an abrupt decline in crater depth beyond D $\sim$~16 km (Ri = 320 m). The 6 km conductive lid over warm convective ice leads to d-D that is well within observations. The 7 km lid is a reasonably good fit, although it is more consistent with the deepest observed craters of any given size.
Overall, the conductive-convective scenario produces good fits for warm convective ice at both temperatures investigated here. The conductive ice lid thickness is 5 km for warm convective ice at 255 K, and 6 - 7 km for warm convective ice at 265 K. The two of the best fits for Ri = 230 m are included in the Supplementary Material (Animations S2 and S3). 
\subsection*{3.2 Comparison to Observed Crater Morphologies}
Figure 9 shows the modeled crater profiles for the three best fit scenarios: the conductive (8 km shell over ocean) and conductive-convective scenarios (5 km ice lid over warm convective ice at 255 K and 6 km ice lid over warm convective ice at 265 K). Figure 1 shows the Europan crater topographic profiles and images for 5 craters ranging from 1.3 km to 23 km in diameter. The crater profiles for panels (a-d) were digitized from Figure 9 in Bray et al. [2014], and panel (e) from Figure 9c in Schenk and Turtle [2009]. Although there are $\sim$~150 catalogued craters over 1 km in diameter on Europa, there is a relatively small number of craters larger than 10 km in diameter [Schenk, 2002; Schenk and Turtle, 2009]. Excluding the multiring basins, only 11 of those have been assigned the morphological type [Schenk and Turtle, 2009]. These craters are listed in Table S4. Since only a handful of topographical profiles are available for comparison to the modeled craters, in this section we perform a qualitative assessment and comparison of modeled and observed craters. 
\subsection*{3.2.1 Simple-to-Complex Transition}
The onset of complex structures occurs at crater diameters of approximately 4 km, which is consistent with observations [Schenk, 2002]. The modeled craters produced by impactors with Ri = 150 m (D $<$ $\sim$~8 km) for all scenarios exhibit complex morphologies, including a central peak. The smallest craters (Ri = 26 m, D $<$ 2 km) show the characteristics consistent with simple craters across all simulations. Beyond D $\sim$~ 4 km (Ri = 70 m), the modeled craters transition into complex structures; however, this transition is more gradual (flat floors, then central peaks) for some cases and more abrupt for others (e.g., Figure 9b,c). In the conductive shell case, the presence of the underlying ocean at depths $>$ 7 km does not have an influence on the simple-to-complex transition, also noted by Bray et al. [2014]. However, the simple-to-complex transition does seem to be a function of thermal gradient. For example, a crater formed by the impactor with Ri = 107 m into an 8 km or a 9 km conductive shell over ocean exhibits a flat floor (Figure 9b). Conversely, the same impact into a warmer target (4 km conductive lid over warm convective ice at 255 K) leads to a complex crater with a well-defined central peak. The three best fits from our simulations are consistent with the morphological simple-to-complex transition (Transition I).
\subsection*{3.2.2 Anomalous Morphologies}
Compared to craters on Ganyamede and Callisto, Europan craters transition to anomalously shallow depths and anomalous morphologies at smaller diameters. This trend is linked to the presence of underlying global ocean at depths of a few tens of kilometers [Pappaardo et al., 1999; Kievelson et al., 2000]. However, complex Europan craters exhibit various morphologies at comparable diameters, making the validation of numerical models to observations challenging. 
For example, Grainne (D $\sim$~14 km) and Amergin (D $\sim$~19 km) are both classified as disrupted central peak craters. Within this size range, there are also the flat floored craters Math (D $\sim$~15 km) and Rhiannon (D $\sim$~16 km), and the central peak craters Eochaird (D $\sim$~17 km) and Cilix (D $\sim$~19 km, Figure 1c). At larger sizes, despite having nearly same diameters, Maeve (D $\sim$~22 km, Figure 2d) and Mannanán (D $\sim$~23 km, Figure 2e) have strikingly different morphologies. Maeve is a classic example of a central peak crater, while Mannanán is a disrupted central peak crater with a central pit [Moore et al., 2001]. 
The modeled craters with D $\sim$~11-12 km (Ri = 230 m) are consistent with the observed classic central peak craters of similar size, such as Avagddu (D $\sim$~11 km). 
\par The morphology of larger modeled craters (D $>$ 15 km), however, varies depending on the setup (conductive versus conductive-convective). For all shell thicknesses modeled in this study, a simple conductive shell over ocean leads to craters with prominent peaks and central pits (Figure 9e,f), consistent with some of the observed craters, such as Maeve and Cilix (Figure 2c,d). However, the formation of central pits might be a byproduct of vertical impact setup and axial symmetry, rather a real effect (see Section 3.3). The disrupted central peaks and flat floors are not well reproduced with the fully convective ice shell.
\par In the conductive ice lid over warm convective ice setup, the modeled craters with D > 15 km exhibit a wider range of morphologies, but generally lack large peaks (Figure 9e,f). The 6 km conductive lid over warm convective ice at 265 K scenario leads to anomalously shallow craters (Figure 9e,f) and craters with irregular crater floor (Figure 9e), reminiscent of a disrupted central peak crater (e.g. Manannán). The largest modeled crater (D $\sim$~22 km, Ri = 405 m) is broadly consistent with the largest shallowest observed crater with the disrupted central peak (Manannán) (Figure 9c). The 5 km conductive lid over warm convective ice at 255 K scenario produces anomalously shallow craters and anomalous morphologies (e.g., flat floors, absence of well-defined crater rims, and irregular crater floors, reminiscent of disrupted central peak). However, the largest modeled crater (D $\sim$~21 km, Ri = 405 m) is significantly deeper than the two observed shallowest large craters on Europa (Figure 9b). It is also deeper than its modeled counterparts using the 8 km conductive shell and the 6 km conductive lid over warm convective ice at 265 K (Figure 9f).    
\subsection*{3.3 Material Splashing and Overflow}
At larger crater diameters, during crater collapse, a central uplift will form. This central uplift is unstable and will also succumb to gravitational collapse. If the thermal gradient of the target is sufficiently high, as it is the case in several scenarios examined in this study (conductive ice lid over warm convective ice), collapse of the central uplift may cause warm material to splash out of the crater center and spread over the crater rim [Johnson et al., 2016]. This occurs for two reasons. First, the combination of a thin conductive lid and a higher thermal gradient contributes to the presence of warm material at a shallow depth. This warm material is also much weaker than the cold material, because the strength of material depends on temperature [Collins et al., 2004], and deforms more readily than cold brittle ice. Second, the axial symmetry setup in hydrocode simulations leads to exaggerated central uplift as opposed to three dimensional models of oblique impacts [Elbeshousen et al., 2009]. Specific simulations affected with the material overflow are those for Ri = 320 m and 405 m impacting a 4, 5 and 6 km conductive lid over warm convective ice at 255 K, and Ri = 405 m impacting a 5 km conductive lid over warm convective ice at 265 K. 
\par The example of this is shown in Figure 10 for a 320 m projectile impacting a 4 km thick ice lid overlying warm convective ice at 255 K. The pre-impact surface and transient crater are shown in panels (a) and (b). During the crater collapse, central uplift forms, as shown in panel (c) at 110 seconds into the simulation. However, due to axial symmetry, the central uplift is exaggerated, as shown in the subsequent time segment in panel (d), at 180 s. Considering that the strength of material depends on temperature [Collins et al., 2004], this warm, weaker material will readily relax (panel (e)) and free-fall into the center of the crater (panel (g)), thereby pushing even more warm material out. The crater rim becomes disguised by a layer of warm material, which will obscure the true dimensions of the crater. Consequently, the new, apparent rim will not correspond to the actual crater rim, which remains concealed. Panels (g) and (h) show the intermediate step shortly before the final shape of the crater takes place at 685 seconds (panel (ii)). 
\par The apparent rim produced by the outflow is broad and inconsistent with the sharp scarps that define observed crater rims (Figure 2). Moreover, the final crater depth will also be affected due to the artificially raised rims such that the craters appear unrealistically deep. On a d-D plot, these craters exhibit depths that increase with an increasing crater diameter, which is not consistent with the observed craters on Europa [Schenk, 2002]. The most extreme example of this is the scenario with the 4 km lid over warm convective ice at 255 K. For example, the crater depth for Ri = 230 m is 0.32 km (not affected by overflow). For comparison, for larger impacts where the crater depth is expected to further decrease, the overflow results in the apparent depth of 0.35 km (Ri = 320 m), and 0.40 km (Ri = 405 m). Animation S4 shows the overflow for Ri = 405 m impacting the 5 km thick conductive ice lid over warm convective ice at 265 K. Data points for simulations that suffer from the apparent warm material overflow are represented with triangles in Figure 8. Finally, we note that larger the impact, more of the overflow effect there will be.  
\par Thus, on one hand, if the warm material overflow is dominated by the axial symmetry setup rather than the behavior imposed by the conditions of the modeled ice shell, then the real depth and diameter dimensions will remain poorly constrained and such results cannot be used to infer the ice shell conditions on Europa. On the other hand, if the axal symmetry plays only a minor role, then none of the scenarios producing skewed d-D (e.g., 4 km ice shell over warm conductive ice at 255 K, see Figure 8b) would be an appropriate consideration for the modeled ice shell conditions on Europa. Alternatively, there could be a more complex interplay between axial symmetry and modeled ice shell; however, more comprehensive studies are needed before a more definitive assertion can be made.  
For other simulations, where the overflow is of lesser extent, we performed the measurement at the point at which the rim fully formed, but before it was flooded over by the warm material (and thus before the entire crater fully formed). The measurement was performed following the same methodology as described earlier in this Section. Therefore, we are able to report and plot (Figure 8b,c) crater dimensions. These are considered special cases and as such, are denoted with half-filled squares in Figure 8(b,c). 
\par An example of the crater profile at two time steps used to measure the dimensions are shown in Figure 11. The profiles represent the results for Ri = 320 m into a 5 km thick ice lid over warm convective ice at 255 K. Since the final crater depth is shallow, and to better show the intermediate and final crater shapes, the profiles are plotted with the exaggerated vertical scale (Figure 11a). The crater rim was measured 140 s into the simulation, as this is when it fully formed, although the crater continues to collapse. The crater depth, on the other hand, was measured at 500 s, when the crater reached its final dimensions. Although minor oscillations (a remnant of splashing) localized at the very narrow central region of the crater floor took longer to completely cease (see Animation S4), they did not affect the final crater morphology or apparent depth and diameter. By t = 500 s, point the rim is no longer raised, but appears flat (panel (c)), with only the inner ‘edge’ present. However, it is evident that the inner edge at 500 s overlaps the location of the rim at 140 s, thus confirming that our approach to measure the crater diameter at earlier time is robust. 

\section*{4. Discussion}
Recent numerical studies [Bray et al., 2014; Cox and Bauer, 2015] reproduced the observed crater d-D and inflection points using a fully conductive icy shell over ocean. Bray et al. [2014] also matched crater morphology (e.g., summit pits and central domes), to discern the thickness of Europa’s ice shell. While numerical modeling using layering can produce domes and summit pits in larger craters (D $>$ 15 km) (see Bray et al. [2014] for more details), there are certain considerations pertaining to hydrocode simulations that need to be examined.

\subsection*{4.1 Numerical Modeling Considerations}
In our simulations, the impactor composition considered was ice, but if it was rocky or metallic, for example, impact-generated heating would be greater [e.g., Barr and Citron, 2011]. In many hydrocodes, including iSALE-2D, the impact angle is set at 90 degrees. In reality, however, the majority of impacts occur at an angle of 45 degrees [Gilbert, 1983; Shoemaker, 1962]. Thus, in some cases, axial symmetry in modeling might not accurately represent the real conditions involved in cratering process. Some of the factors that may be sensitive to angle of impact are the cratering efficiency [Elbeshausen et al., 2009], impact generated heating and melting [Pierazzo and Melosh, 2000], exaggerated central uplift, and formation of central features (e.g., pits, domes and summit pits). Additionally, the effect of impact velocity cannot be neglected, as it affects the extent of melting [Pierazzo et al., 1997] and final crater morphology [Silber et al., 2017].  We will briefly touch on the effect of impact angle and velocity in this section.

\subsection*{4.1.1 The effect of impact angle}
Cratering efficiency decreases as a function of impact angle, and transient crater depth, diameter and volume all depend on impact angle [Elbeshausen et al., 2009]. Vertical impacts also produce a symmetrical isobaric core and the melting region around the point of impact [Pierazzo and Melosh, 2000]. As the angle of impact decreases from the vertical (90 deg), the strength of the shock and the shape of the heated region changes accordingly and becomes more and more asymmetrical. Vertical impacts generate substantially more heating and melt as opposed to very oblique impacts ($<$ 30 deg), where the melting efficiency is greatly diminished [Pierazzo and Melosh, 2000]. These varying conditions might influence how central features, specifically pits and domes, form (e.g., formation of melt pools and conditions conductive to post-impact drainage). Moreover, axial symmetry often leads to exaggerated central uplift when compared to full 3D simulations of oblique impacts (see Results). Subsequent relaxation of the uplifted material could alter the final morphology [Moore et al., 2017] of the crater center by producing certain central features which might not form in oblique impacts (e.g., in a 3D simulation). Hence, the formation of central features, such as pits, domes and summit pits in a crater might be sensitive to the angle of impact and form mainly through the axial symmetry, as opposed to using full three-dimensional modeling of oblique impacts [Elbeshausen et al., 2009]. However, since there are no studies that specifically focused on this problem, it is not possible to make a conclusive determination as to what extent these features might or might not be affected by the impact angle. While the height of a central peak might be affected by axial symmetry, a study on Venusian craters has shown that the central peak diameter and its position relative to crater center has no correlation to the impact angle, at least for impacts on rocky bodies [Ekholm and Melosh, 2001].  

\subsection*{4.1.2 The effect of impact velocity}
Impact velocity also plays an important factor in cratering process. For example, impacts at high velocities generate substantially more melt than impacts at low velocities [Pierazzo et al., 1997], which is important for creating conditions conducive to formation of melt pools. The average impact velocity on Europa is 26 km/s [Zanhle et al., 2003], higher than that implemented in this study and other numerical studies [e.g., Bray et al., 2014; Cox and Bauer, 2015]. Moreover, the morphology (e.g., onset of flat floors and central peaks) of impact craters, especially in the simple-to-complex transition regime is highly sensitive to impact velocity, as shown by recent study on lunar craters [Silber et al., 2017]. Silber et al. [2017] modeled lunar craters (D = 10 – 26 km) forming as a result of impacts at velocities 6 – 15 km/s. Modeling at very low impact velocities (2 – 10 km/s), representative of conditions at Pluto, also shows variations in crater depth and simple-to-complex transition as a function of impact velocity [Bray et al., 2015]. Given the small statistics of catalogued Europan craters with D $>$ 10 km and their apparent diverse morphologies even at similar diameters (see Table S4), it is not possible to deduce which factors outlined above might dominate the observed morphologies. Moreover, at larger crater sizes, in addition to variations in morphology at the same crater sizes (see Results and Table S4), there is large ambiguity in whether central features form during crater collapse or well after.   
If central pits and domes form post impact, then depending on the extent of the melt pool size, volume and its emplacement within ice, the post-impact modification could result in a variety of possible crater morphologies. The conductive-convective scenario (thick ice) would be most consistent with such hypothesis. However, if central pits and domes form during the cratering collapse, then the purely conductive ice shell over ocean (thin shell) would be the most appropriate choice.  

\subsection*{4.2 The effect of strength of ice on crater size and morphology}
The strength of ice (Y) as a function of depth is an important consideration and as demonstrated through our simulations, multiple strength profiles as a function of depth are possible (Figure 4a). Ultimately, it is the parameters for strength of ice that dictate what combination of thermal gradient, roll over temperature, and conductive lid thickness will be required to produce a given strength profile. The strength parameters for ice are derived [Bray et al., 2014 and references therein] from experimental data [Beeman, 1988], and the strength model used in iSALE is described in detail in Ivanov et al. [1997] and Collins et al. [2004]. Therefore, we will only discuss the outcome in the context of our results. The strength-depth profiles for our best fits, represented with solid lines in Figure 4a, illuminate the linkage between crater morphology and the strength of ice. For example, the 8 km conductive ice shell over the ocean is notably stronger down to a depth of ~ 7 km than either of the best fits for the conductive lid over warm convective ice. However, below 8 km, it shifts to the strengthless regime. The strength profile for the conductive ice lid over warm convective ice is highly sensitive to the temperature of convecting ice. The 6 km conductive ice over warm convective ice at 265 K is stronger than the 5 km conductive ice lid over warm convective ice at 255 K down to depths of ~ 6 km, when the trend reverses and the strength becomes notably stronger for the cooler ice (Figure 4a). This might be the reason why a conductive layer needs to be thinner in order to produce the same depth crater when the lower, ductile ice layer is cooler and therefore stronger. Moreover, this may also explain why the conductive ice lid over warm convective ice scenario cannot produce the central peaks as large as the purely conductive ice shell over ocean. Although our results are robust for the best fit strength-depth profiles, they strongly depend on previously established parameters for ice based on laboratory experiments [Beeman, 1988]. Any future improvements to ice parameters would also invoke an adjustment to thermal gradient, roll over temperature, and possibly conductive lid thickness.

\subsection*{4.3 Formation of pits and domes: A post-impact process?}
\par Moore et al. [2017] reinforced the hypothesis that central pits and domes develop after the impact, suggesting that the morphology of the mature impact structure is strongly dependent on the melt pool size, volume and its emplacement within ice. In principle, the water lens will undergo upwelling as it freezes out, altering the final crater morphology. Hence, the central pit and dome formation during the crater collapse does not have to be invoked if the crater d-D agree with observations. As noted by Moore et al. [2017], impactor composition and velocity will also affect the extent of post-impact formation of domes and pits. Schenk [2002] reported that all catalogued craters on Europa are unmodified and unrelaxed suggesting that post impact formation of pits and domes proposed by Moore et al. [2017] may not affect crater depths.
\par Although the 8 km conductive ice shell over ocean and the 6 km conductive ice lid over warm convective ice scenarios lead to the very similar d-D trend, the latter is more conducive to formation of pits and domes [Moore et al., 2017] due to higher thermal gradient and impact induced heating on a larger scale (Figure 7). Schenk [2002] interpreted Transition II on Europa as the expression of temperature-dependent rheologic change (e.g., warm convective ice) at the depths of 7 – 8 km, also possibly representing the thickness of a stagnant non-convecting lid [McKinnon, 1999]. Indeed, our results for a 5 km and a 6 – 7 km thick conductive lid overlying warm convective ice at 255 K and 265 K, respectively, are consistent with Schenk’s [2002] interpretation. 
\par Post-impact modification [e.g., Moore et al., 2017] might better explain a larger spectrum of diverse and intermixed morphologies observed on Europa (Table S4). In context of our results, we prefer the conductive-convective model (i.e., the conductive ice lid overlying warm convective ice (with ocean at some depth) scenario), as it is more consistent with the notion that pits and domes form post impact, the variety of Europan morphologies at larger crater sizes, as well as the hypothesis indicating that Transition III is symptomatic of underlying ocean at depths 19 – 25 km [Schenk, 2002]. Despite these remarks, it cannot be conclusively established whether Europa’s ice shell is purely conductive (thin shell), or if it has a thick layer of warm convective ice beneath a conductive lid (thick shell).
\subsection*{4.4 Implications for Future Work}
To evaluate and test the validity of the hypothesis that central pits and domes form post-impact, more focused studies at higher resolution (to evaluate production of melt), with different impactor compositions, and representative impact velocities at Europa are recommended. While modeling at high impact velocities is computationally more demanding and expensive, future studies should also investigate morphological differences arising from impacts at varying velocities into icy bodies, such as Europa. 
Test runs with larger impactors (Ri $>$ 405 m) confirm relationships shown by Cox and Bauer [2015] that there is either a breach (ice shell over ocean) or the final crater becomes extremely shallow (nearly flat) and thus neither depth or diameter can be confidently discerned. While the resolutions used in our study are not sufficient to resolve any faulting indicative of rings formation, this might correspond to the progression to Transition III [Schenk 2002]. The ring structures should be very sensitive to the difference in rheology between warm convecting ice and liquid water [Singer et al., 2013]. Thus, modeling the formation of multiring basin on Europa may be provide important information about the structure of the ice shell and could be sensitive to ocean thickness even in the case of a conductive lid overlying warm convecting ice. The aim of future work will be to model these largest craters.

\section*{5. Conclusions}
In our study, we modeled formation of impact craters on Europa to investigate thickness and internal structure of its ice shell. Our modeling results suggest that both a fully conductive shell (thin shell) and conductive lid over warm convective ice (thick shell) are capable of reproducing the observed crater morphologies and depth-diameter on Europa. If the ice shell is indeed thick, our study places an estimate on the conductive portion of the crust to 5 – 7 km, depending on the temperature of warm convecting ice. In particular, the ice shell structures that best match the observations are: a 5 km thick conductive ice lid overlying warm convective ice at 255 K, a 6 – 7 km thick conductive ice lid overlying warm convective ice at 265 K, and an 8 km completely conductive ice shell over ocean. The latter is consistent with previous studies that placed the ice shell thickness between 7 and 10 km [Bray et al., 2014; Cox and Bauer, 2015]. If central features within the crater, such as pits and domes, form during crater collapse, our simulations are in better agreement with a conductive shell (thin shell). If central features form well after the impact, however, our simulations are more consistent with a conductive-convective shell (thick shell).    

\section*{Acknowledgements}
The authors thank Gareth Collins, Ronadh Cox and one anonymous reviewer for their helpful comments to improve this paper. EAS gratefully acknowledges the Natural Sciences and Engineering Research Council of Canada (NSERC) Postdoctoral Fellowship program for supporting this project. We gratefully acknowledge the developers of iSALE-2D (www.isale-code.de), the simulation code used in our research, including Gareth Collins, Kai Wünnermann, Dirk Elbeshausen, Boris Ivanov and Jay Melosh. All data associated with this study are listed tables in the Supplemental Material and shown in figures. The simulations were performed using iSALE r-1967. The simulation outputs are available upon request from elizabeth\textunderscore silber@brown.edu. The simulation inputs and model outputs are published on Dataverse (doi:10.7910/DVN/EY6MNT). 

\section*{References}
Amsden, A. A., H. M. Ruppel, and C. W. Hirt (1980), SALE: A Simplified ALE computer program for fluid flow at all speeds, Los Alamos National Laboratories Report, LA-8095(June), 101p-101p.
\newline Barr, A. C., and R. I. Citron (2011), Scaling of melt production in hypervelocity impacts from high-resolution numerical simulations, Icarus, 211(1), 913-916, doi:10.1016/j.icarus.2010.10.022.
\newline Beeman, M., W. B. Durham, and S. H. Kirby (1988), Friction of ice, Journal of Geophysical Research: Solid Earth, 93(B7), 7625-7633, doi:10.1029/JB093iB07p07625.
\newline Bray, V. J., G. S. Collins, J. V. Morgan, H. J. Melosh, and P. M. Schenk (2014), Hydrocode simulation of Ganymede and Europa cratering trends – How thick is Europa’s crust?, Icarus, 231, 394-406, doi:10.1016/j.icarus.2013.12.009.
\newline Bray, V. J., G. S. Collins, J. V. Morgan, and P. M. Schenk (2008), The effect of target properties on crater morphology: Comparison of central peak craters on the Moon and Ganymede, MAPS, 43(12), 1979-1992, doi:10.1111/j.1945-5100.2008.tb00656.x.
\newline Collins, G. S., H. J. Melosh, and B. A. Ivanov (2004), Modeling damage and deformation in impact simulations, MAPS, 39(2), 217-231, doi:10.1111/j.1945-5100.2004.tb00337.x.
\newline Cox, R., and A. W. Bauer (2015), Impact breaching of Europa's ice: Constraints from numerical modeling, Journal of Geophysical Research: Planets, 120(10), 1708-1719, doi:10.1002/2015JE004877.
\newline Cox, R., L. C. F. Ong, M. Arakawa, and K. C. Scheider (2008), Impact penetration of Europa's ice crust as a mechanism for formation of chaos terrain, MAPS, 43(12), 2027-2048, doi:10.1111/j.1945-5100.2008.tb00659.x.
\newline Ekholm, A. G., and H. J. Melosh (2001), Crater features diagnostic of oblique impacts: The size and position of the central peak, Geophysical Research Letters, 28(4), 623-626, doi:10.1029/2000gl011989.
\newline Elbeshausen, D., K. Wünnemann, and G. S. Collins (2009), Scaling of oblique impacts in frictional targets: Implications for crater size and formation mechanisms, Icarus, 204(2), 716-731, doi:10.1016/j.icarus.2009.07.018.
\newline Greenberg, R., P. Geissler, B. R. Tufts, and G. V. Hoppa (2000), Habitability of Europa's crust: The role of tidal-tectonic processes, Journal of Geophysical Research: Planets, 105(E7), 17551-17562.
\newline Hoppa, G. V. (1999), Formation of Cycloidal Features on Europa, Science, 285, 1899-1902.
\newline Ivanov, B., D. Deniem, and G. Neukum (1997), Implementation of dynamic strength models into 2D hydrocodes: Applications for atmospheric breakup and impact cratering, International Journal of Impact Engineering, 20(1), 411-430.
\newline Ivanov, B. A., F. Langenhorst, A. Deutsch, and U. Hornemann (2002), How strong was impact-induced CO2 degassing in the Cretaceous-Tertiary event? Numerical modeling of shock recovery experiments, in Catastrophic events and mass extinctions: impacts and beyond, edited by C. Koeberl and K. G. MacLeod, Geological Society of America.
\newline Ivanov, B. A., H. J. Melosh, and E. Pierazzo (2010), Basin-forming impacts: Reconnaissance modeling, edited by R. L. Gibson and W. U. Reimold, Geological Society of America.
\newline Johnson, B. C., et al. (2016a), Formation of the Orientale lunar multiring basin, Science, 354(6311), 441-444, doi:10.1126/science.aag0518.
\newline Johnson, B. C., T. J. Bowling, A. J. Trowbridge, and A. M. Freed (2016b), Formation of the Sputnik Planum basin and the thickness of Pluto's subsurface ocean, Geophysical Research Letters, 43(19), 10,068-010,077, doi:10.1002/2016GL070694.
\newline Khurana, K. K., M. G. Kivelson, D. J. Stevenson, G. Schubert, C. T. Russell, R. J. Walker, and C. Polanskey (1998), Induced magnetic fields as evidence for subsurface oceans in Europa and Callisto, Nature, 395(6704), 777-780.
\newline Kivelson, M. G. (2000), Galileo Magnetometer Measurements: A Stronger Case for a Subsurface Ocean at Europa, Science, 289, 1340-1343, doi:10.1126/science.289.5483.1340.
\newline Kivelson, M. G., K. K. Khurana, D. J. Stevenson, L. Bennett, S. Joy, C. T. Russell, R. J. Walker, C. Zimmer, and C. Polanskey (1999), Europa and Callisto: Induced or intrinsic fields in a periodically varying plasma environment, Journal of Geophysical Research: Space Physics, 104(A3), 4609-4625, doi:10.1029/1998JA900095.
\newline McKinnon, W. B. (1999), Convective instability in Europa's floating ice shell, Geophysical Research Letters, 26(7), 951-954, doi:10.1029/1999GL900125.
\newline Melosh, H. J. (1979), Acoustic fluidization: A new geologic process?, Journal of Geophysical Research: Solid Earth, 84(B13), 7513-7520, doi:10.1029/JB084iB13p07513.
\newline Melosh, H. J. (1989), Impact Cratering - A Geologic Process.
\newline Melosh, H. J., A. G. Ekholm, A. P. Showman, and R. D. Lorenz (2004), The temperature of Europa's subsurface water ocean, Icarus, 168(2), 498-502, doi:10.1016/j.icarus.2003.11.026.
\newline Melosh, H. J., E. V. Ryan, and E. Asphaug (1992), Dynamic fragmentation in impacts: Hydrocode simulation of laboratory impacts, Journal of Geophysical Research: Planets, 97(E9), 14735-14759, doi:10.1029/92JE01632.
\newline Moore, J. M., et al. (2001), Impact Features on Europa: Results of the Galileo Europa Mission (GEM), Icarus, 151(1), 93-111, doi:10.1006/icar.2000.6558.
\newline Moore, J. M., et al. (1998), Large Impact Features on Europa: Results of the Galileo Nominal Mission, Icarus, 135(1), 127-145, doi:10.1006/icar.1998.5973.
\newline Moore, J. M., P. M. Schenk, and D. G. Korycansky (2017), Large impact features on icy Galilean satellites, paper presented at Lunar and Planetary Science XLVIII, The Woodlands, TX.
\newline Nimmo, F., and M. Manga (2009), Geodynamics of Europa’s icy shell, in Europa, edited, pp. 382-404, The University of Arizona Press USA.
\newline Ojakangas, G. W., and D. J. Stevenson (1989), Thermal state of an ice shell on Europa, Icarus, 81(2), 220-241, doi:10.1016/0019-1035(89)90052-3.
\newline Pappalardo, R. T., et al. (1999), Does Europa have a subsurface ocean? Evaluation of the geological evidence, Journal of Geophysical Research: Planets, 104(E10), 24015-24055, doi:10.1029/1998JE000628.
\newline Pierazzo, E., and H. J. Melosh (2000), Melt Production in Oblique Impacts, Icarus, 145(1), 252-261, doi:10.1006/icar.1999.6332.
\newline Pike, R. J. (1980), Control of crater morphology by gravity and target type - Mars, Earth, Moon, paper presented at Lunar and Planetary Science Conference 11th Proceedings, New York, Pergamon Press, Houston, TX, March 17-21, 1980.
\newline Prockter, L. M., and R. T. Pappalardo (2000), Folds on Europa: Implications for Crustal Cycling and Accommodation of Extension, Science, 289(5481), 941-943, doi:10.1126/science.289.5481.941.
\newline Ruiz, J., L. Montoya, V. López, and R. Amils (2007), Thermal Diapirism and the Habitability of the Icy Shellof Europa, Origins of Life and Evolution of Biospheres, 37(3), 287-295.
\newline Ruiz, J., and R. Tejero (2000), Heat flows through the ice lithosphere of Europa, Journal of Geophysical Research, 105(E12), 29283-29283, doi:10.1029/1999JE001228.
\newline Schenk, P. M. (2002), Thickness constraints on the icy shells of the galilean satellites from a comparison of crater shapes, Nature, 417(6887), 419-421.
\newline Schenk, P. M., and E. P. Turtle (2009), Europa’s impact craters: Probes of the icy shell, in Europa, edited by W. B. M. R. T. \newline Pappalardo, K. K. Khurana, pp. 181-198, University of Arizona Press, Tuscon, Arizona.
\newline Schmidt, B. E., D. D. Blankenship, G. W. Patterson, and P. M. Schenk (2011), Active formation of ‘chaos terrain’ over shallow subsurface water on Europa, Nature, 479, 502-505, doi:10.1038/nature10608.
Senft, L. E., and S. T. Stewart (2011), Modeling the morphological diversity of impact craters on icy satellites, Icarus, 214(1), 67-81, doi:10.1016/j.icarus.2011.04.015.
\newline Silber, E. A., G. R. Osinski, B. C. Johnson, and R. A. F. Grieve (2017), Effect of impact velocity and acoustic fluidization on the simple-to-complex transition of lunar craters, Journal of Geophysical Research: Planets, 122(5), 800-821, doi:10.1002/2016JE005236.
\newline Singer, K., W. McKinnon, and P. Schenk (2013), Ice lithosphere thickness on Europa from impact basin ring-graben, paper presented at Lunar and Planetary Science Conference.
\newline Thompson, S. L., and H. S. Lauson (1972), Improvements in the Chart-D Radiationhydrodynamic Code III: Revised Analytic Equations of State Rep., Sandia National Laboratories.
\newline Tillotson, J. H. (1962), Metallic equations of state for hypervelocity impactRep., 141 pp, DTIC Document.
Turtle, E. P., and E. Pierazzo (2001), Thickness of a Europan Ice Shell from Impact Crater Simulations, Science, 294(5545), 1326-1328, doi:10.1126/science.1062492.
\newline Wünnemann, K., G. S. Collins, and H. J. Melosh (2006), A strain-based porosity model for use in hydrocode simulations of impacts and implications for transient crater growth in porous targets, Icarus, 180(2), 514-527, doi:10.1016/j.icarus.2005.10.013.
\newline Zahnle, K., P. Schenk, H. Levison, and L. Dones (2003), Cratering rates in the outer Solar System, Icarus, 163(2), 263-289, doi:10.1016/s0019-1035(03)00048-4.

\newpage

\section*{Figures}

\begin{figure}[ht] 
\includegraphics[width=0.8\textwidth]{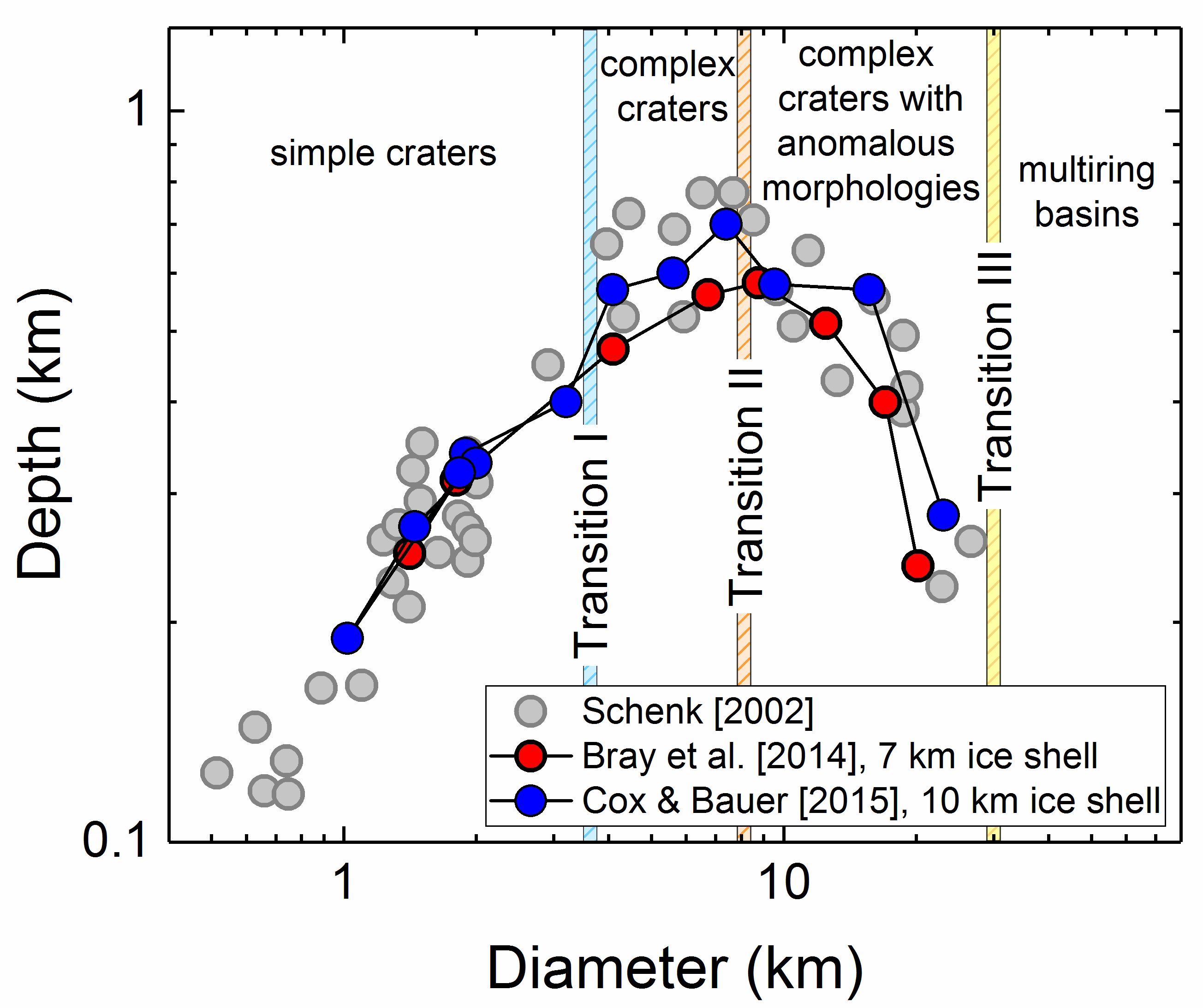}

\caption{\color{Gray} The depth-diameter (d-D) for modeled craters as given by Bray et al. [2014] and Cox and Bauer [2015], plotted against the observed d-D [Schenk 2002]. Bray et al. [2014] placed the conductive shell thickness at 7 km, while the estimate by Cox and Bauer [2015] was 10 km. In our study, in addition to ice shell over ocean, we also consider the conductive ice over warm convective ice scenario. The three transitions as defined by [Schenk 2002] are annotated on the plot.}
\end{figure}

\newpage
\begin{figure}[ht] 

\includegraphics[width=\textwidth,height=0.8\textheight,keepaspectratio]{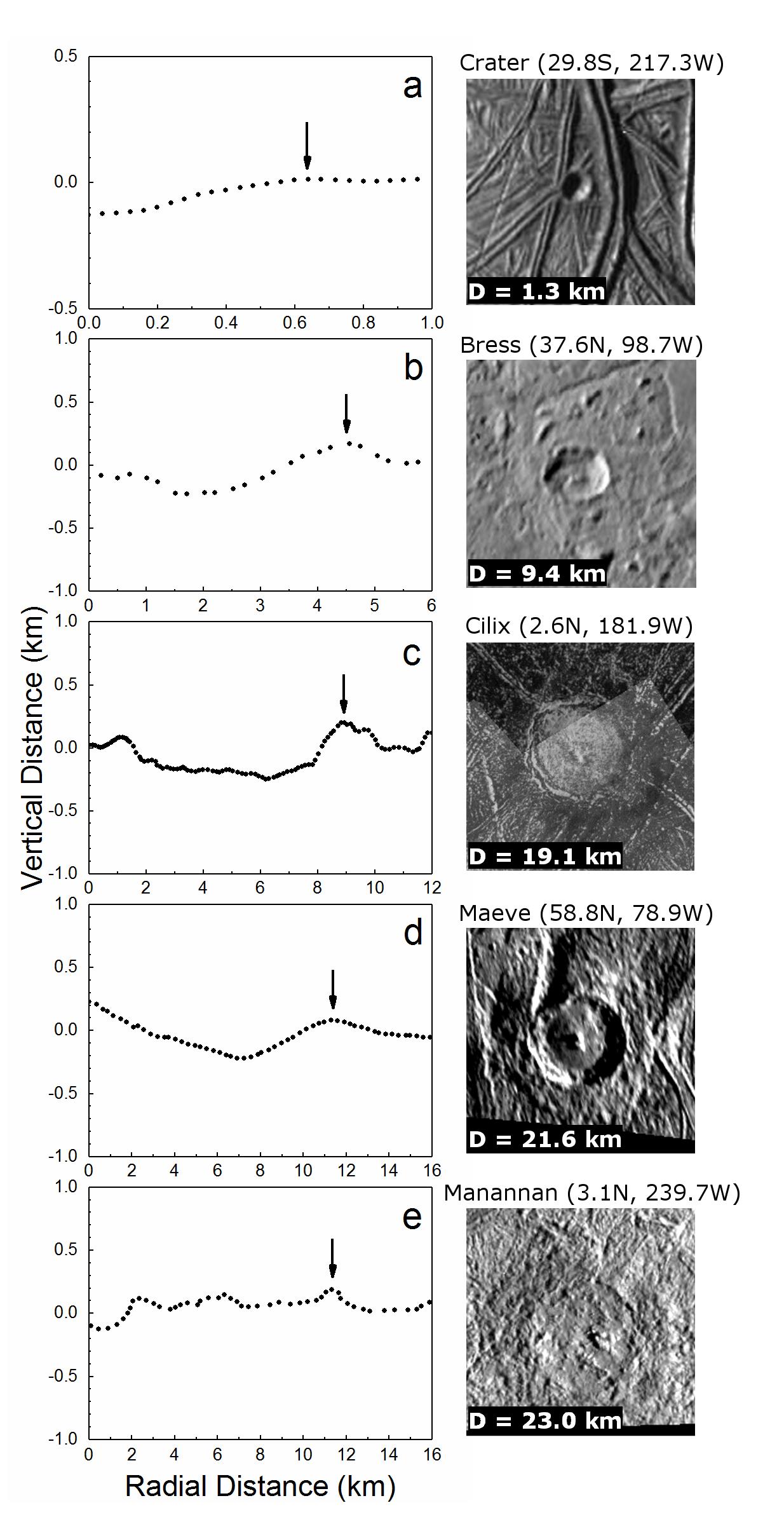}

\caption{\color{Gray} The Europan crater profiles (left column) alongside the crater images (right column). The crater rims are shown with arrows. All panels, except (a), have the same vertical scale. The horizontal scale varies across panels. The crater profiles for panels (a-d) were digitized from Figure 6 in Bray et al. [2014], and panel (e) from Figure 9c in Schenk and Turtle [2009]. Crater images are from the Galileo spacecraft. All craters, except (a), are complex craters with anomalous morphologies (Transition II) [Schenk 2002].}
\end{figure}

\newpage
\clearpage
\begin{figure}[ht] 
\includegraphics[width=0.8\textwidth]{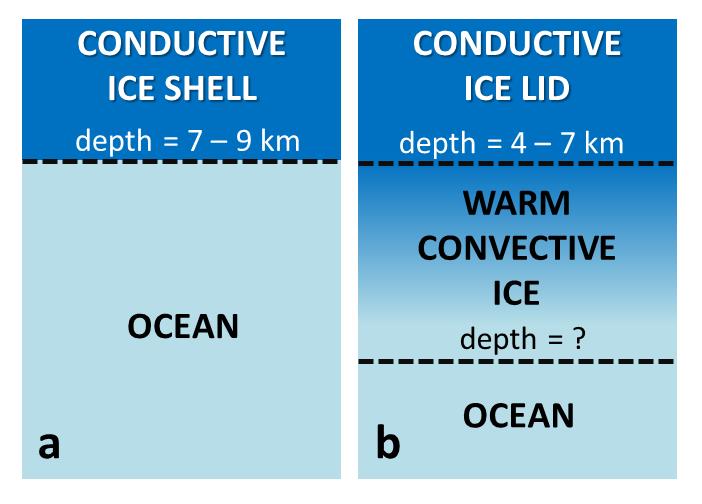}

\caption{\color{Gray} The diagram depicting the two setups modeled in our study: (a) a fully conductive shell over ocean, and (b) a conductive ice lid overlying warm convective ice. Schenk [2002] estimated the depth of the ocean at 19-25 km and noted that craters at Transition III would exhibit features indicative of the rheological changes (e.g. multiring structures). While it is assumed that there is ocean at some depth in the conductive-convective scenario (panel (b)), it is not explicitly modeled. This is because at crater sizes investigated in this study (D $<$ 23 km), the presence of the ocean at depths proposed by Schenk [2002] is not expected to make an appreciable contribution.}
\end{figure}

\newpage
\begin{figure}[ht] 
\includegraphics[width=\textwidth,height=0.65\textheight,keepaspectratio]{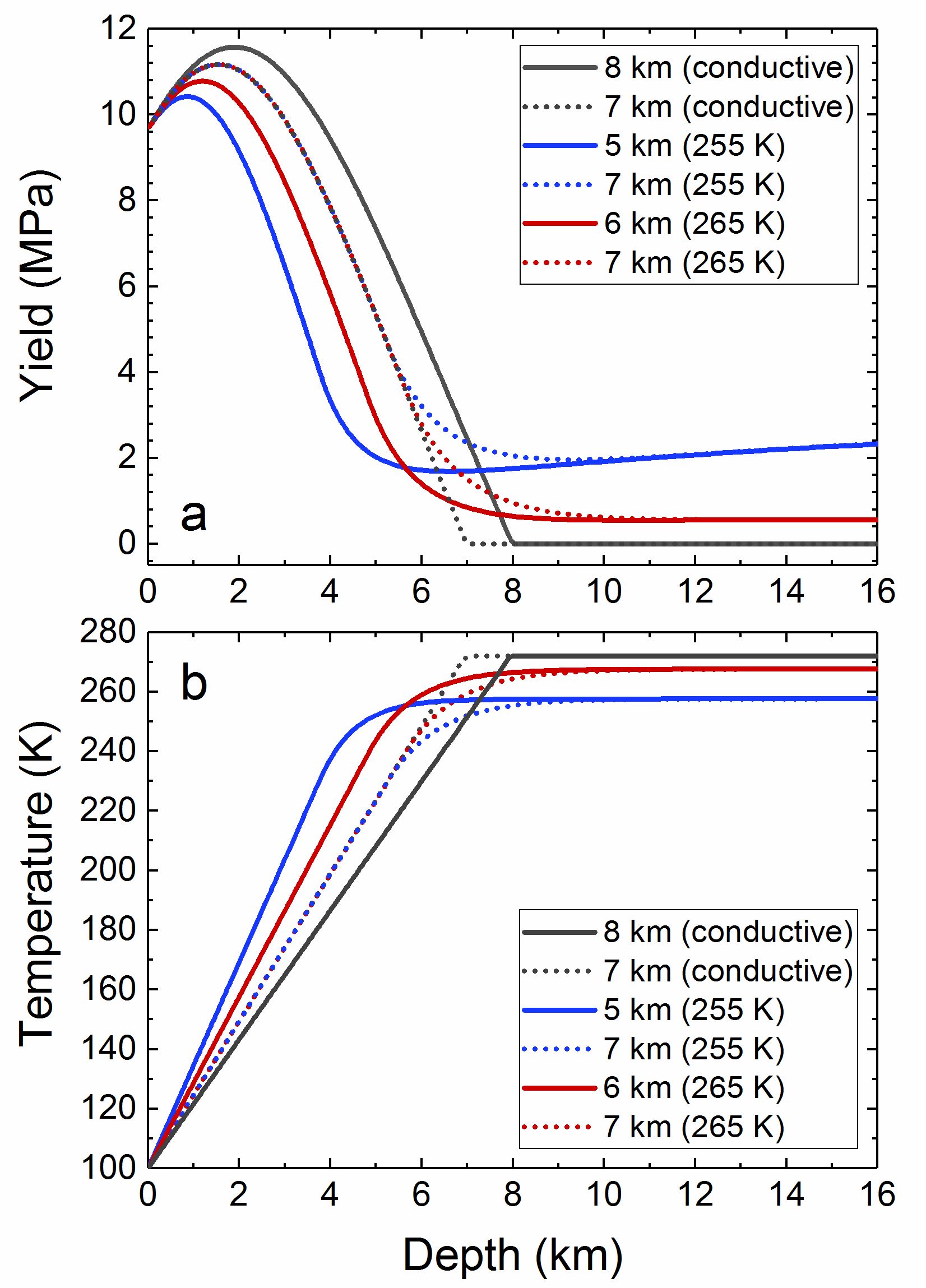}

\caption{\color{Gray} Examples of (a) strength-depth and (b) temperature-depth profiles used in our simulations. Legend reports conductive lid thickness and temperature of the warm convecting ice (e.g., solid red curve is 6 km thick conductive layer overlying warm ice at 265 K).}
\end{figure}

\newpage
\begin{figure}[ht] 
\includegraphics[width=\textwidth,height=0.6\textheight,keepaspectratio]{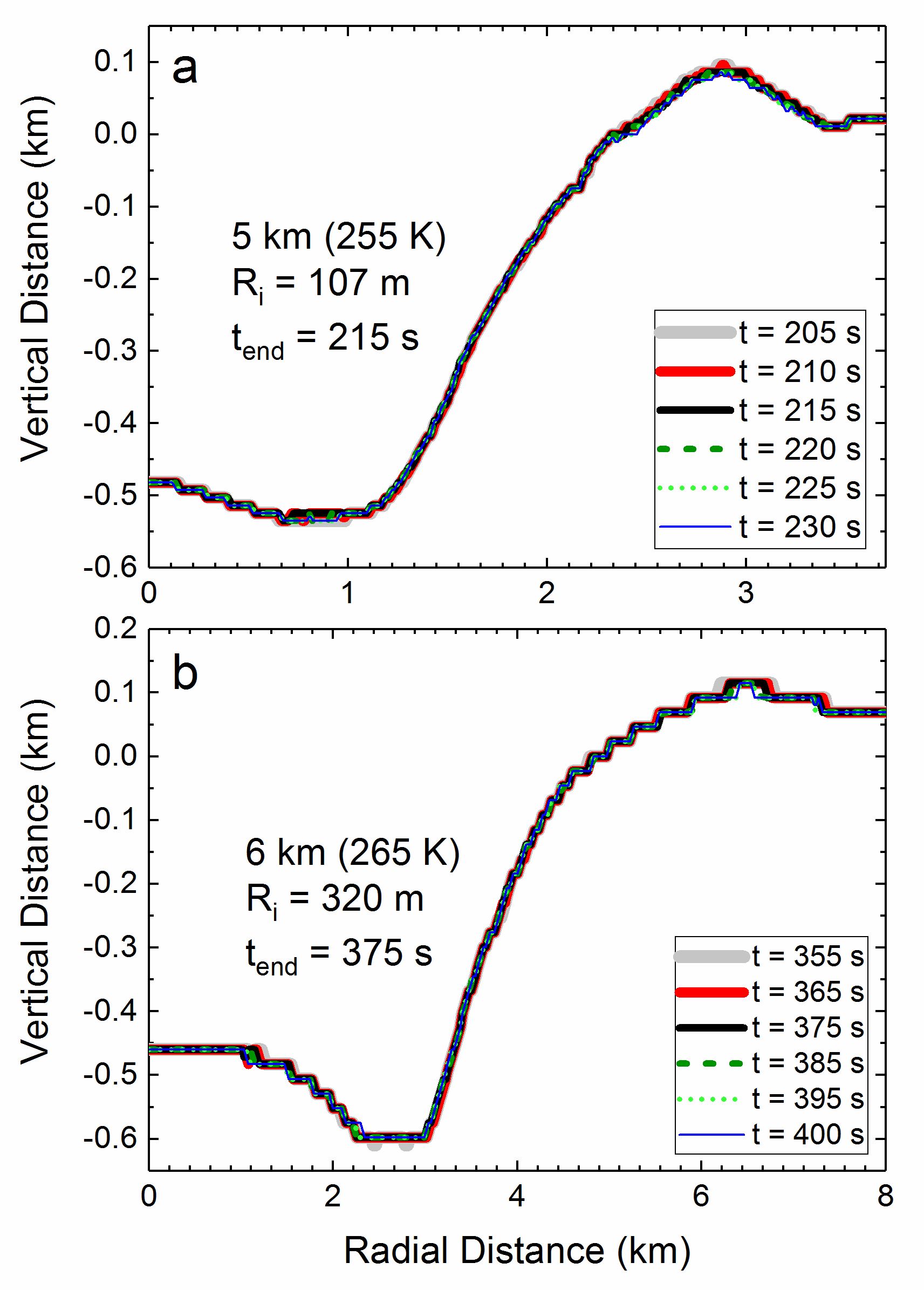}

\caption{\color{Gray} Examples of crater profiles used to determine when the crater fully formed for two representative simulations. The end times for each simulation are reported in Tables S1-S3. The simulation is considered over when crater collapse is finished, and no notable change in crater depth, diameter and morphology takes place.  The simulation scenario, impactor radius and reported end time ($t_{end}$) are shown in figure panels. Note that the vertical and horizontal scales in the two panels are different. The times prior to ($t_{end}$) are denoted with solid grey and red lines, and the later times with dark green and light green broken lines, and solid blue line (also shown in legend).}
\end{figure}

\newpage
\begin{figure}[ht] 
\includegraphics[width=\textwidth,height=\textheight,keepaspectratio]{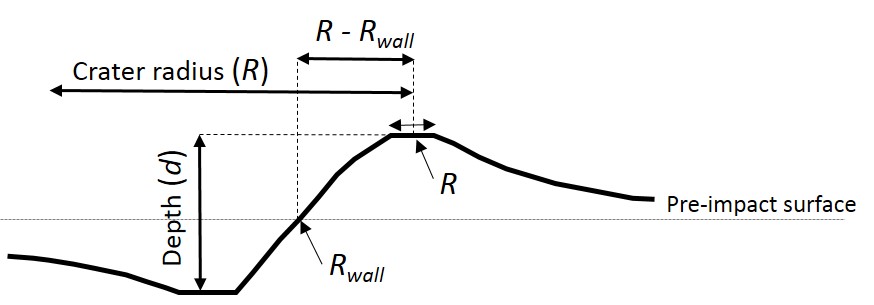}

\caption{\color{Gray} Diagram depicting the method for measuring crater depth and diameter, as well as the uncertainty in crater diameter. Since the highest point on the crater rim can be anywhere from one to several cells wide, depending on a simulation, we first found the median (R), which also serves as the measurement of the crater radius (diameter, D = 2R). Then, we obtained the intersecting point (Rwall) of the crater wall and the pre-impact surface. The uncertainty in crater radius is expressed as Rerr = 0.25(R - Rwall), and uncertainty in overall diameter as Derr = 2Rerr. Note that the diagram is meant for visual purposes only, and as such, it does not represent the actual crater proportions.}
\end{figure}

\newpage
\begin{figure}[ht] 
\includegraphics[width=0.7\textwidth,height=0.6\textheight,keepaspectratio]{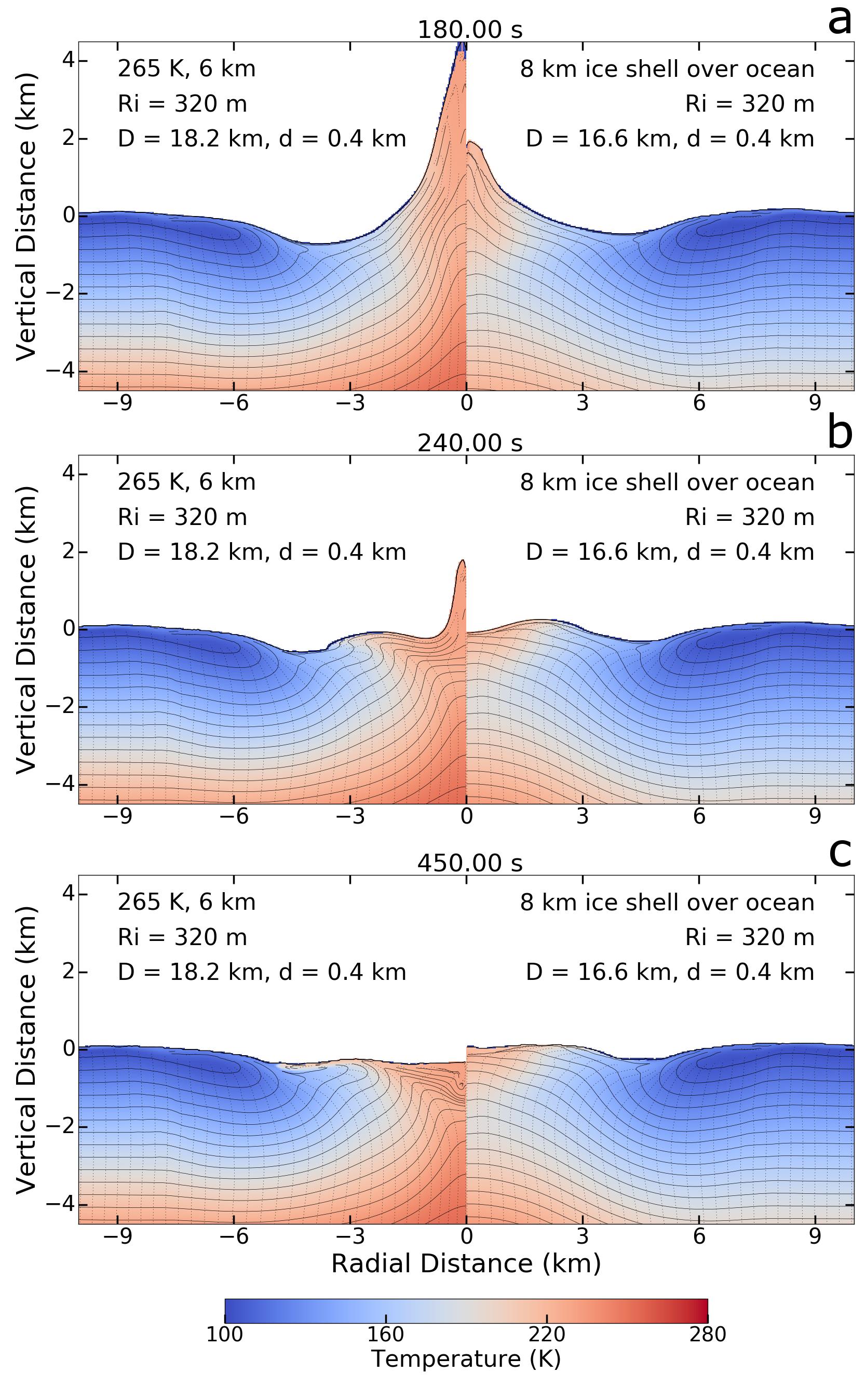}

\caption{\color{Gray} Time series of the craters produced by a 320 m in radius impactor. Shown are the time steps at (a) 180 s, (b) 240 s, and (c) 440 s. In all panels, the 6 km thick conductive lid over warm convective ice at 265 K (left) and the 8 km ice shell over ocean scenario (right) are plotted side by side for better comparison. In ice over ocean case, the uplifted crater floor is due to the contribution of the ocean underneath. There is a larger uplift of much warmer and more ductile material (see Section 3.3) in the convective ice case as compared to the 8 km ice shell at 180 s and 240 s into the simulation (Figure 4a,b).}
\end{figure}

\newpage
\begin{figure}[ht] 
\includegraphics[width=\textwidth,height=0.6\textheight,keepaspectratio]{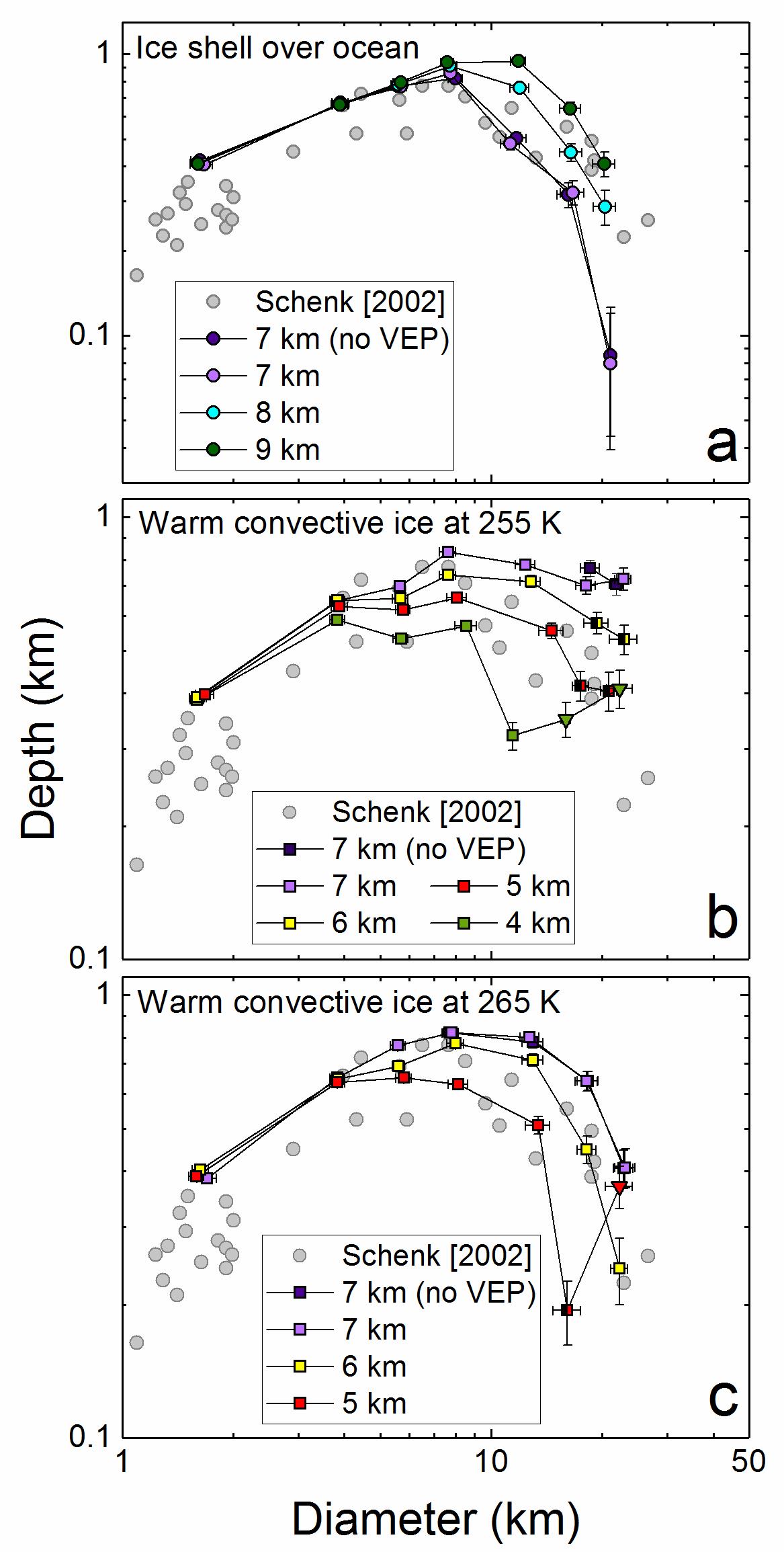}

\caption{\color{Gray} Crater depth versus diameter for the (a) ice shell over ocean, and warm convective ice at (b) 255 K and (c) 265 K scenarios, plotted against observed d-D [Schenk, 2002]. Note that the vertical scale in (a) differs from that in (b) and (c). Triangles represent craters for which it was not possible to reliably measure d-D due to warm material overflow, and half-filled points represent the craters for which it was possible to determine the location of the crater rim before it was engulfed by the warm material (see Section 3.3). While these points are not listed separately in legend, their color corresponds to the simulation setups shown in (b) and (c).}
\end{figure}

\newpage
\begin{figure}[ht] 
\includegraphics[width=\textwidth,height=0.6\textheight,keepaspectratio]{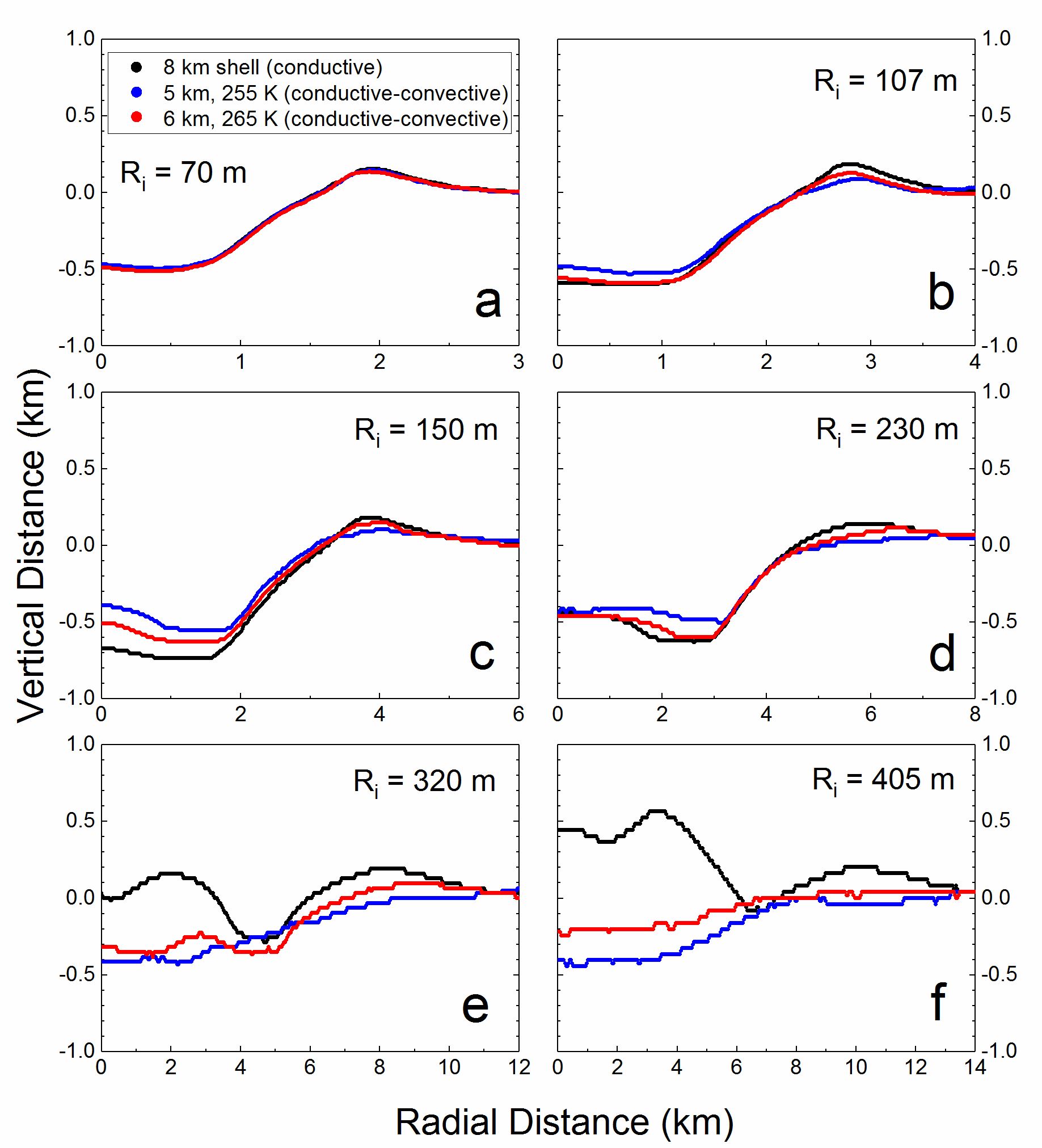}

\caption{\color{Gray} The modeled crater profiles for best fit results from our simulations for conductive scenario (8 km shell over ocean (black)) and conductive-convective scenarios (5 km ice lid over warm convective ice at 255 K (blue), and 6 km ice lid over warm convective ice at 265 K (red)). The legend shown in panel (a) applies to all panels.}
\end{figure}

\newpage
\begin{figure}[ht] 
\includegraphics[width=\textwidth,height=0.7\textheight,keepaspectratio]{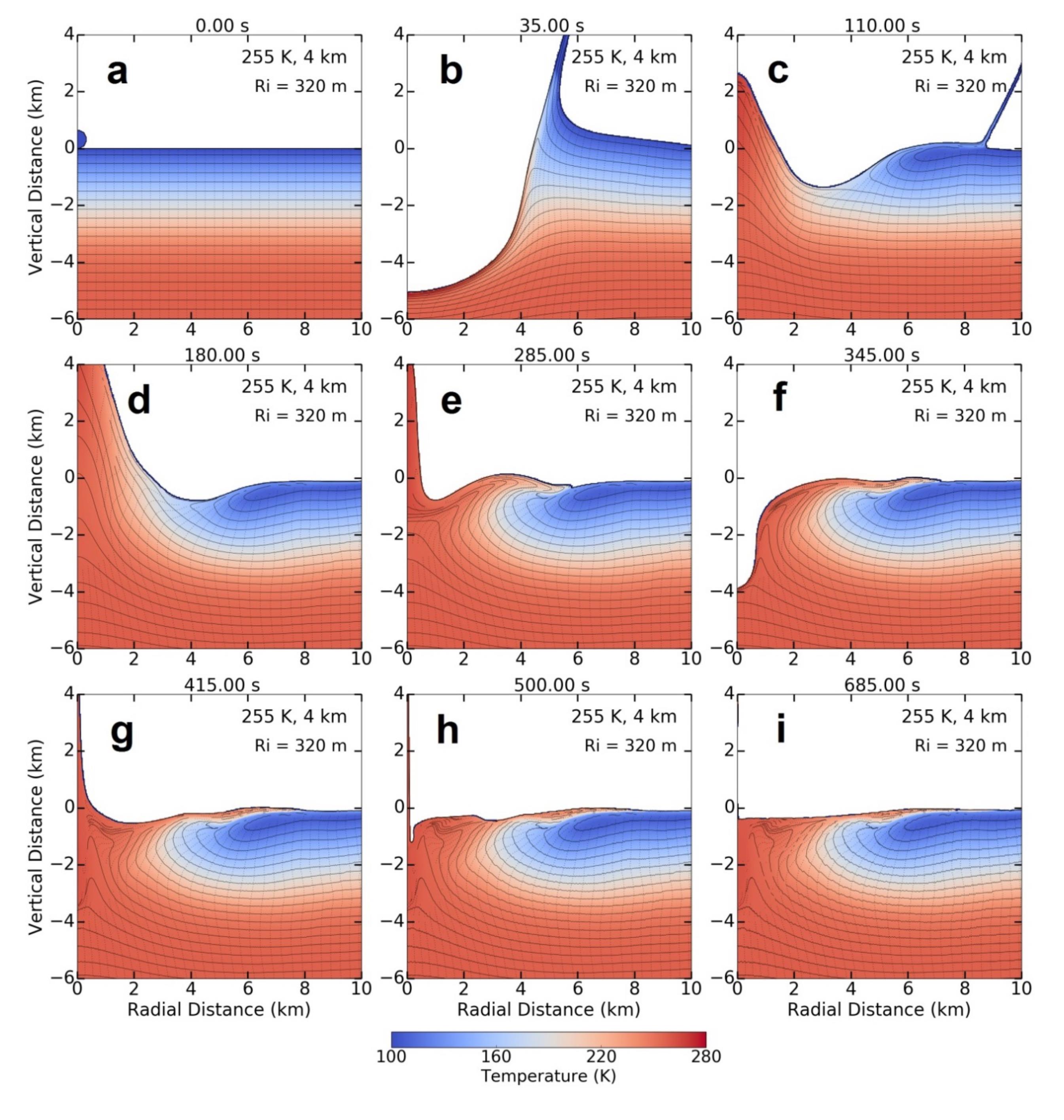}

\caption{\color{Gray} Time series showing the crater formation for the projectile with Ri = 320 m impacting a 4 km thick ice lid over warm convective ice at 255 K.}
\end{figure}

\newpage
\begin{figure}[ht] 
\includegraphics[width=\textwidth,height=0.7\textheight,keepaspectratio]{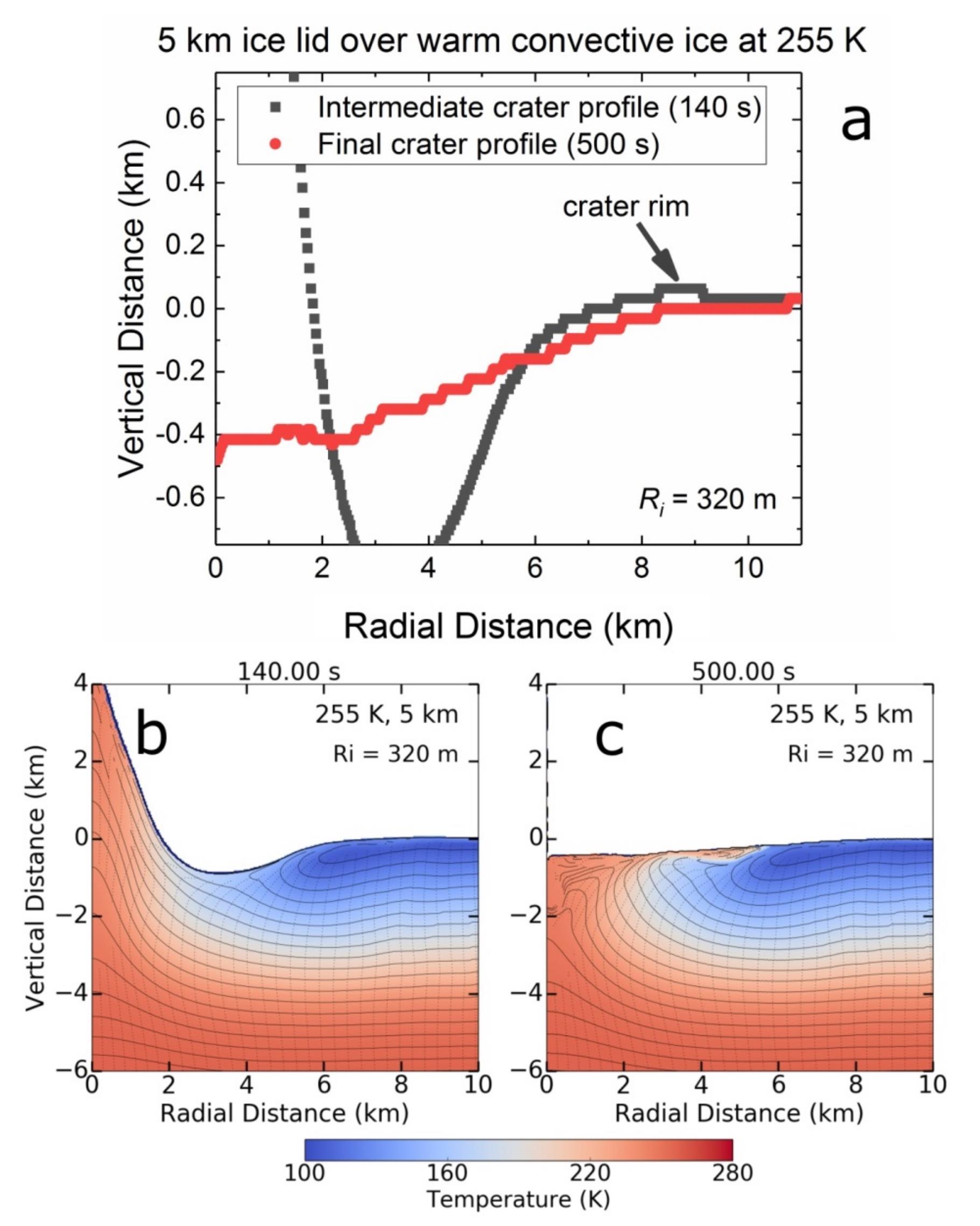}

\caption{\color{Gray} (a) The crater profiles at 140 s and 500 s into the simulation. Panels (b) and (c) show the time series at 140 s and 500 s. Note that the vertical scale in (b,c) is different from that in (a).}
\end{figure}


\newpage
\section*{Tables.} 

\setcounter{figure}{0}
\begin{table}[ht] 
\includegraphics[width=0.6\textwidth,height=0.7\textheight,keepaspectratio]{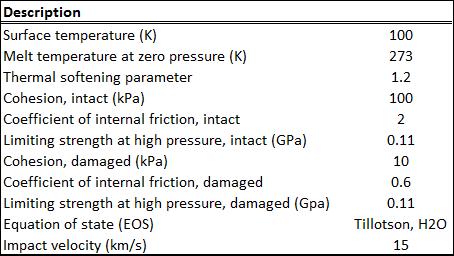}
\caption{\color{Gray} (a) The crater profiles at 140 s and 500 s into the simulation. Panels (b) and (c) show the time series at 140 s and 500 s. Note that the vertical scale in (b,c) is different from that in (a).}
\end{table}

\newpage
\begin{table}[ht] 
\includegraphics[width=0.4\textwidth,height=0.7\textheight,keepaspectratio]{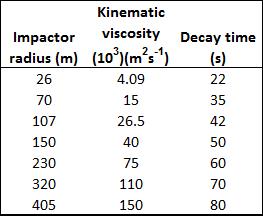}

\caption{\color{Gray} (a) The crater profiles at 140 s and 500 s into the simulation. Panels (b) and (c) show the time series at 140 s and 500 s. Note that the vertical scale in (b,c) is different from that in (a).}
\end{table}

\newpage
\begin{table}[ht] 
\includegraphics[width=0.5\textwidth,height=0.7\textheight,keepaspectratio]{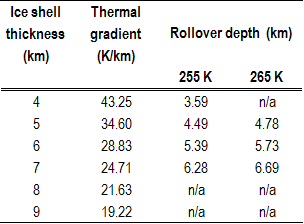}

\caption{\color{Gray} (a) The crater profiles at 140 s and 500 s into the simulation. Panels (b) and (c) show the time series at 140 s and 500 s. Note that the vertical scale in (b,c) is different from that in (a).}
\end{table}

\section*{Supporting Information}
Supporting information is published on line (JGR - Planets). The simulation inputs and model outputs are published on Dataverse (doi:10.7910/DVN/EY6MNT).




\nolinenumbers



\end{document}